\PassOptionsToPackage{dvipsnames}{xcolor}

\documentclass[sigconf,screen,authorversion,nonacm]{acmart} 

\usepackage{fontawesome5} 
\usepackage[nolist,nohyperlinks]{acronym}
\usepackage{colortbl}
\usepackage{multirow}
\usepackage{enumitem} 
\usepackage[dvipsnames]{xcolor}
\usepackage{soul} 
\usepackage[nameinlink]{cleveref}
\usepackage{xspace}
\usepackage{forloop}
\usepackage{calc}
\usepackage{balance} 
\usepackage{hyperref} 

\newif\ifanonymous

\newcommand{\anony}[1]{%
  \ifanonymous
    <Anonymous for submission>%
  \else
    #1%
  \fi
}

\newcounter{loopcntr}
\newcommand{\dashedline}[2]{%
  \noalign{\vskip -5pt} 
  \multicolumn{#1}{c}{%
    \makebox[\linewidth]{%
      \forloop{loopcntr}{0}{\value{loopcntr} < #2}{%
        \rule[0.2ex]{0.2cm}{0.1pt}\hspace{0.11cm}%
      }%
    }%
  }\\
  \noalign{\vskip 1pt} 
}

\newenvironment{commentwrapper}[1]{\color{#1}}{\color{Black}}

\begin{acronym}
\acro{HCI}{Human-Computer Interaction}
\acro{RQ}{research question}
\acro{VR}{Virtual Reality}
\acro{AR}{Augmented Reality}
\acro{MR}{Mixed Reality}
\acro{XR}{Extended Reality}
\acro{VE}{virtual environment}
\acro{AI}{Artificial Intelligence}
\acro{NPC}{non-player character}

\acro{PXI}{Player Experience Inventory}
\acro{IMI}{Intrinsic Motivation Inventory}
\acro{RPE}{Ratings of Perceived Exertion}
\acro{PENS}{Player Experience of Need Satisfaction}
\acro{IoS}{Inclusion of the Other in the Self}
\acro{CCPIG}{Competitive and Cooperative Social Presence in Gaming}
\acro{CPM}{Cooperative Performance Metric}
\acro{PACES}{Physical Activity Enjoyment Scale}
\acro{HR}{heart rate}
\end{acronym}

\newcommand{\sss}[0]{\textit{Space Scavenger Squad}\xspace}

\acrodefplural{VE}[VEs]{virtual environments}
\acrodefplural{NPC}[NPCs]{non-player characters}
\acrodefplural{RQ}[RQs]{research questions}
\acrodefplural{HR}[HRs]{heart rates}
\acrodefplural{PACES}[PACES-S]{short version of the Physical Activity Enjoyment Scale}

\newcommand{\intext}[2]{``\textit{#1}'' (\textbf{\textit{P#2}})}
\newcommand{\p}[1]{\textbf{\textit{P#1}}}

\newcommand{\mean}[1]{\emph{M}$=$#1}
\newcommand{\sd}[1]{\emph{SD}$=$#1}

\newcommand{\F}[2]{\emph{F}(#1)$=$#2}

\newcommand{\ttest}[2]{\emph{t}(#1)$=$#2}
\newcommand{\psmall}{\emph{p}$<$.001}
\newcommand{\pequal}[1]{\emph{p}$=$#1}

\newcommand{\effectpar}[1]{$\eta^{2}_{P}=$#1}

\newcommand{\wil}[1]{\emph{V}$=$#1}
\newcommand{\kend}[1]{\emph{W}$=$#1}

\newcommand{\n}[1]{\emph{n}$=$#1}
\newcommand{\Fman}[2]{{$\chi$}$^2$(#1)$=$#2}

\definecolor{red}{RGB}{255,105,97}
\definecolor{blue}{RGB}{121,163,210}
\definecolor{green}{RGB}{121,210,143}

\definecolor{purple}{RGB}{210, 121, 203}
\definecolor{orange}{RGB}{255,153,102}

\newcommand{\blue}[1]{\textcolor{blue}{#1}}
\newcommand{\green}[1]{\textcolor{green}{#1}}
\newcommand{\purple}[1]{\textcolor{purple}{#1}}
\newcommand{\orange}[1]{\textcolor{orange}{#1}}

\newcommand{\VOne}{\green{\faIcon{clone}}~\textsc{free}\xspace}
\newcommand{\VTwo}{\blue{\faIcon{flickr}}~\textsc{coupled}\xspace}
\newcommand{\VThree}{\purple{\faIcon{cogs}}~\textsc{concurrent}\xspace}

\newcommand{\VOneN}{\rev{\textsc{free}}\xspace}
\newcommand{\VTwoN}{\rev{\textsc{coupled}}\xspace}
\newcommand{\VThreeN}{\rev{\textsc{concurrent}}\xspace}

\newcommand{\VOneQ}{\textsc{free}}
\newcommand{\VTwoQ}{\textsc{coupled}}
\newcommand{\VThreeQ}{\textsc{concurrent}}

\definecolor{blueB}{RGB}{0,0,255}
\newcommand{\rev}[1]{#1}

\AtBeginDocument{%
  \providecommand\BibTeX{{%
    \normalfont B\kern-0.5em{\scshape i\kern-0.25em b}\kern-0.8em\TeX}}}

\anonymousfalse 

\begin{document}
\title[Exploring the Dynamics of Player Cooperation in a Co-located Cooperative Exergame]{From Solo to Social: \rev{Exploring the Dynamics of Player Cooperation in a Co-located Cooperative Exergame}}
\author{Derrick M. Wang}
\authornote{Authors are also affiliated with the Department of Systems Design Engineering, University of Waterloo.}
\authornote{\textbf{This is a pre-print version of our CHI 2025 paper. Please check out the DOI for the published version: \url{https://doi.org/10.1145/3706598.3713937}.}}
\email{dwmaru@uwaterloo.ca}
\orcid{0000-0003-3564-2532}
\affiliation{
   \institution{HCI Games Group, Stratford School of Interaction Design and Business, University of Waterloo}
  \city{Waterloo}
  \state{ON}
  \country{Canada}
}

\author{Sebastian Cmentowski}
\email{s.cmentowski@tue.nl}
\orcid{0000-0003-4555-6187}
\affiliation{
\department{Industrial Design} 
\institution{Eindhoven University of Technology} 
\city{Eindhoven}
\country{Netherlands}
}

\author{Reza Hadi Mogavi}
\email{rhadimog@uwaterloo.ca}
\orcid{0000-0002-4690-2769}
\affiliation{
   \institution{HCI Games Group, Stratford School of Interaction Design and Business, University of Waterloo}
  \city{Waterloo}
  \state{ON}
  \country{Canada}
}

\author{Kaushall Senthil Nathan}
\authornotemark[1]
\email{k3senthi@uwaterloo.ca}
\orcid{0009-0009-0846-1593}
\affiliation{
   \institution{HCI Games Group, Stratford School of Interaction Design and Business, University of Waterloo}
  \city{Waterloo}
  \state{ON}
  \country{Canada}
}

\author{Eugene Kukshinov}
\email{eugene.kukshinov@uwaterloo.ca}
\orcid{0000-0002-3759-5218}
\affiliation{
   \institution{HCI Games Group, Stratford School of Interaction Design and Business, University of Waterloo}
  \city{Waterloo}
  \state{ON}
  \country{Canada}
}

\author{Joseph Tu}
\authornotemark[1]
\email{joseph.tu@uwaterloo.ca}
\orcid{0000-0002-7703-6234}
\affiliation{
   \institution{HCI Games Group, Stratford School of Interaction Design and Business, University of Waterloo}
  \city{Waterloo}
  \state{ON}
  \country{Canada}
}

\author{Lennart E. Nacke}
\email{lennart.nacke@acm.org}
\orcid{0000-0003-4290-8829}
\affiliation{%
     \institution{HCI Games Group, Stratford School of Interaction Design and Business, University of Waterloo}
    \city{Waterloo}
    \state{ON}
    \country{Canada}
}

\renewcommand{\shortauthors}{Wang, et al.}

\begin{abstract}
Digital games offer rich social experiences and promote valuable skills, but they fall short in addressing physical inactivity. Exergames, which combine exercise with gameplay, have the potential to tackle this issue. \rev{However, current exergames are primarily single-player or competitive. To explore the social benefits of cooperative exergaming}, we designed a custom co-located cooperative exergame that features three distinct \rev{forms of cooperation}: \textsc{free} \rev{(baseline)}, \textsc{coupled}, and \textsc{concurrent}. We conducted a within-participants, mixed-methods study ($N=24$) to evaluate these designs and their impact on players' enjoyment, motivation, and performance. Our findings reveal that cooperative play improves social experiences. It drives increased team identification and relatedness. Furthermore, our qualitative findings support cooperative exergame play. This has design implications for creating exergames that effectively address players' exercise and social needs. Our research contributes guidance for developers and researchers who want to create more socially enriching exergame experiences.
\end{abstract}



\begin{teaserfigure}
  \includegraphics[width=\textwidth]{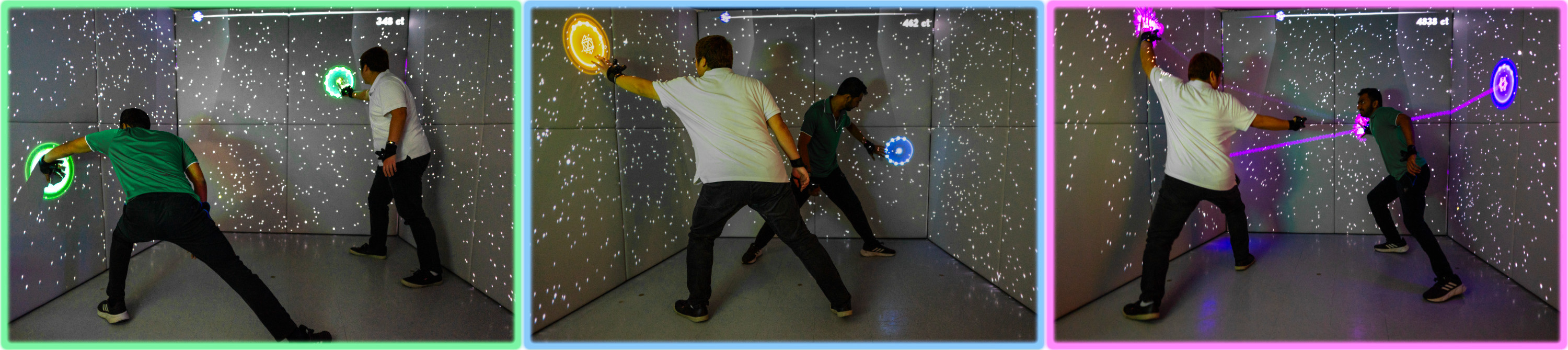}
  \captionsetup{justification=centering}
  \caption{In our cooperative co-located exergame \textit{Space Scavenger Squad}, players work together to catch dots that appear on the ExerCube's projection walls. In a user study, we compared three cooperation mechanics: \VOne, where players share one task; \VTwo, where each player catches differently-colored dots, and \VThree, where players have to sync their actions to succeed.} 
  \Description{This figure presents a visual demonstration of gameplay in our MR-based cooperative exergame Space Scavenger Squad. In the three conditions, players perform dashing movements to physically tap on the panels of the ExerCube to score. You can find detailed descriptions in Section 3.}
  \label{fig:teaser}
\end{teaserfigure}

\maketitle
\section{Introduction}
\label{sec:intro}
Despite lingering stereotypes of gamers as antisocial introverts, digital gaming can be a highly social activity. Far from isolating players, video games often thrive on shared experiences. Whether competing in friendly matches or cooperating in teams, the social aspect of gaming is central to its widespread appeal and cultural impact~\cite{bowman_social_2022}. However, the advantages of multiplayer social play go beyond an improved player experience. The existing literature shows that cooperative games can teach essential social skills by improving team cohesion~\cite{greitemeyer_theres_2013, zheng_impact_2021}, involvement\rev{~\cite{reuter_game_2014, cheng_haptic_2014}}, or communication\rev{~\cite{nasir_cooperative_2013, hernandez_design_2014}}. Given these social benefits, it is no surprise that digital games are increasingly adopted for team-building activities in various settings~\cite{ellis_games_2008, keith_team_2018,  alharthi_investigating_2021}. Consequently, they possess similar qualities as physical group play in the context of cooperative teamwork---both fulfill our need for social connections and interactions\rev{~\cite{kaos_social_2019, moreno_socially_2013}}.

However, there is arguably one significant shortcoming of existing digital games when comparing them to traditional sports: they do not benefit physical health. This aspect is crucial, given the ``sedentary epidemic'' (i.e., most people do not meet the recommended levels of physical activity~\cite{statistics_canada_almost_2017, elgaddal_physical_2022}.) Notably, there is one game genre aimed towards addressing this pressing societal problem: exergames. These games~\cite{mueller_exertion_2016}, which encourage players to engage in physical exercise by integrating it into the core gameplay, have even been shown to offer a more enjoyable and less exerting experience than traditional physical exercises~\cite{yu_effects_2020, ramirez-granizo_effect_2020, martin-niedecken_exercube_2019}. Using exergames to improve physical health or deliver engaging rehabilitation interventions has, therefore, been explored by a growing body of research, with recent studies particularly focusing on immersive applications~\cite{karaosmanoglu_born_2024}. However, most of these exergames are purely single-player experiences~\cite{chan_social_2024, peng_playing_2013, chan_motivational_2019} or feature competitive modes~\cite{shaw_competition_2016, song_effects_2013}. This focus on individual success ignores the valuable social benefits of engaging in shared physical play and working towards a common goal.

\rev{Previous research on cooperative exergaming has provided some indications of its benefits with regards to motivation~\cite{marker_better_2015, peng_playing_2013, lee_psychological_2017,cheng_haptic_2014}, enjoyment~\cite{peng_playing_2013, deng_which_2024,marquez_segura_design_2013}, and performance~\cite{deng_which_2024, martin-niedecken_towards_2019}}. However, these studies mainly focused on adding an additional player without exploring the role of a specifically-tailored cooperative game design towards facilitating social interactions. For example, \citet{peng_playing_2013} compared the performances of two players playing an exergame together and playing the same game alone. \citet{martin-niedecken_towards_2019} explored custom rules to make a single-player exergame playable by pairs (e.g., one player carrying the other piggyback). Accordingly, the social interactions within these studies are mainly a product of the participants' own creativity and tendency for social behaviour. This is not a result of cooperative game designs and, therefore, may not be generalizable for other exergames or different players~\cite{chan_social_2024}. \rev{Therefore, the potential of cooperative exergames for both physical well-being and social connections~\cite{hamalainen_utilizing_2015, martin-niedecken_towards_2019} remains severely underexplored.} 

\rev{To create an exergaming experience that facilitates and encourages social play, we first need to understand how cooperative game mechanics foster social interactions in exergames.} Cooperation can come in various forms, and different game mechanics may have different effects on the players and may not work in the context of co-located exergames. However, to date, no research has yet explored how mechanics \rev{tailored to cooperation can} affect the player experience in cooperative exergames, specifically with regards to enjoyment, motivation, and performance. Also, we must understand how players cooperate in such games to ultimately learn which concepts are best suited to achieve both socially enriching and physically beneficial exergames. To address this critical gap in previous research, our work focused on two central research questions:

\begin{enumerate}[label= \textbf{RQ\arabic*:}]
    \item What impact do different cooperation designs have on enjoyment, motivation, and performance in a co-located cooperative exergame?
    \item How do players cooperate in a co-located cooperative exergame?
\end{enumerate}

As our first step towards answering these questions, we drew from a recent framework of cooperative game mechanics by \citet{pais_living_2024} to develop a custom cooperative, co-located exergame \rev{(see detailed selection process in \autoref{sec:mechanics})}. As we aim to create an exergame that offers both rich social play and exergame play, it is important to strike a balance between the two~\cite{ghoshal_co-located_2022}. Therefore, we selected the \ac{MR}-based ExerCube platform to host our exergame. The shared physical space and tangible displays allowed us to leverage more immersive gameplay and the physical co-location for players to engage in a fulfilling exergaming experience with rich social interactions~\cite{martin-niedecken_towards_2019}. Our prototype features \rev{three game designs featuring various degrees of cooperation} that allow us to gain a good understanding of their effect on the exergame experience:
\begin{itemize}
    \item \textbf{Free cooperation \rev{(baseline)}:} The game \rev{presents the same task to both players, which makes cooperation possible but not mandatory. Since this design is similar to that of existing work, it serves as our} baseline for understanding individual players' movement patterns and cooperation strategies.
    \item \textbf{Coupled cooperation:} The game presents player-specific targets \rev{spread across the entire play space}. \rev{This is a spatial coordination gameplay mechanic that requires players to navigate the shared space.}
    \item \textbf{Concurrent cooperation:} The game requires players to tap two connected orbs simultaneously. \rev{This is a temporal coordination gameplay mechanic that demands players to coordinate and synchronize their actions.}
\end{itemize}

After finalizing our prototype, we conducted a within-participants study with 24 participants (in 12 dyads). We assessed both quantitative and qualitative measures to gather a comprehensive understanding of their exergaming experience, preferences, and feedback. The results \rev{showed that our cooperative designs led to} improved Relatedness, Team Identification, and exercise performance \rev{compared to the baseline where cooperation is optional}. Our study also revealed the pivotal role social interaction and teamwork play in creating a positive feedback loop that enhances players' exergaming enjoyment, motivation, and performance. Additionally, we examined aspects of our exergame's design, including the gameplay, the narrative, and the audiovisual design, with the participants, and condensed these insights into design implications and recommendations to further improve the cooperative play in future exergames.

Our research proposes three main contributions to the field of \ac{HCI} and Game Research. First, we present an exergame artifact, developed through multiple design cycles and evaluated with empirical data, for which our findings suggest that it offers an enjoyable, motivating, and effective exergaming experience. Second, we establish precedence in research by focusing on cooperation designs within the unique context of co-located cooperative exergames. Finally, we build a foundation for future social exergame design and research based on the insights gained from the empirical data in our study.

\section{Related Work}
\label{sec:relatedwork}
Our research explores players' cooperation in co-located exergames and its impact on player enjoyment, motivation, and performance. In this section, we summarize the relevant previous research.

\subsection{Cooperative Game Design/Mechanics}
Cooperative game mechanics require players to work together rather than against each other~\cite{reuter_game_2014}. In previous literature~\cite{beznosyk_influence_2012, zheng_impact_2021, jin_when_2017, nasir_cooperative_2013}, this term is sometimes used interchangeably with collaborative game mechanics, whereas other researchers distinguish the two with contextual factors. For example, \citet{baek_comparing_2017} distinguish cooperation and collaboration in game-based learning by the role the teacher plays. \citet{reuter_game_2014} define cooperation as temporary alliances and collaboration as more permanent. Following this connotation, we chose to refer to players working together consistently as cooperation, because this arrangement is usually temporary by nature (i.e., until the game ends).

In their recently published framework, \citet{pais_living_2024} have built upon classic definitions from \citet{guimaraes_rocha_game_2008} and \citet{reuter_game_2014} to categorize cooperative mechanics and provide guidelines for future design. \rev{This framework categorizes cooperative games in four different aspects: \textit{Play Structures}, \textit{Player Context}, \textit{Forms of Cooperation}, and \textit{Cooperation Design Patterns}. We selected the \textit{Forms of Cooperation} category as the basis for describing and discussing our exergame design. As this category focuses on distinguishing mechanics based on the different kinds of in-game cooperation they promote, it aligns well with our research goals of evaluating player cooperation as an outcome of tailored cooperative designs. We also considered other categories of this framework but found them less relevant or not offering strong distinctions among our conditions (e.g., the \textit{Player Context} category describes relationships between players; our conditions are consistent in that regard since players interact as teammates with each other and share the same game world and viewpoints.)}

Within this category, the authors identified three subcategories:

\begin{enumerate}
     \item \textbf{Arrangement:} How cooperative tasks are assigned to players (i.e., free, coupled, strict, coincident). 
     \item \textbf{Synchronicity:} How cooperative tasks are done in terms of timing (i.e., sequential, concurrent, asynchronous). 
     \item \textbf{Communication:} How communication is required by the game (i.e., agnostic, limited, required/incentivized). 
 \end{enumerate}

\subsubsection{Effects of Cooperative Play}
Cooperative play has been proven to enhance player experience in various contexts, including massively multiplayer online role-playing games (MMORPG)~\cite{lim_computer_2010, poor_collaboration_2014}, co-located multiplayer games (couch co-op)~\cite{emmerich_impact_2017, garcia_promoting_2022}, gamification~\cite{baek_comparing_2017}, and networked remote exergaming~\cite{mueller_design_2008}. Research has shown that cooperation fosters: (1) enhanced enjoyment and motivation~\cite{marker_better_2015, peng_influence_2012, daggubati_effect_2016, deng_which_2024}, (2) improved communication and social skills\rev{~\cite{bowman_social_2022, hernandez_design_2014}}, and (3) development of spatial awareness and problem-solving skills~\cite{bowman_social_2022, chaarani_association_2022}. Given that cooperative play thrives in both sports and video gaming~\cite{marker_better_2015, engels_effects_2020}, exergames, which combine elements of both, may also benefit from these cooperative designs by enhancing physical and social well-being.

The familiarity between players may influence cooperative behaviours and experience in games as well. So far, research has not been able to reach a consensus on this topic. On the one hand, studies from \citet{hudson_familiarity_2015, mason_friends_2013} and \citet{ravaja_spatial_2006} suggest that playing with friends results in higher trust, social presence, and performance, than playing with strangers in team-based games. On the other hand, studies from \citet{vella_impact_2017} and \citet{karaosmanoglu_playing_2023} suggest that familiarity makes no significant impact on players' social and player experiences. Notably, some authors argue that their findings may be limited to the genre of the games evaluated~\cite{hudson_familiarity_2015} and the type of interactions the games fostered~\cite{karaosmanoglu_playing_2023}. Therefore, we believe that our research can also further expand on these findings by exploring how different cooperative mechanics influence the social experience. 

\subsection{Exergaming}
Exergames (and a number of synonymous terms such as \textit{exertion games}~\cite{mueller_exertion_2016, marshall_expanding_2016}, \textit{active video games}~\cite{kooiman_active_2018, monedero_energy_2017}, \textit{motion-based games}~\cite{gerling_full-body_2012}) are digital games with tasks requiring players to exert physical effort to complete~\cite{mueller_exertion_2016}. In some literature, this is also referred to as ``bodily play''~\cite{mueller_designing_2017}. Examples of popular commercial exergames include \textit{Wii Fit}~\cite{nintendo_wii_2007}, \textit{Dance Dance Revolution}~\cite{konami_dance_1998}, and \textit{Ring Fit Adventure}~\cite{hiroshi_matsunaga_ring_2019}. Research has reported various health benefits similar to regular workouts and a higher enjoyment~\cite{kooiman_active_2018, yu_effects_2020, hadi_mogavi_jade_2024}. This is achieved by directing players' attention from the exertion and the repetitive nature of physical exercise to the immersive and enjoyable experience of video gaming~\cite{lee_psychological_2017, feltz_exergames_2019, ketelhut_integrating_2022, faric_what_2019}.

\subsubsection{Immersive exergaming and the ExerCube}
\label{sec:exercube}
Immersive exergaming has quickly gained popularity following breakthroughs in immersive technology (i.e., \ac{XR}, including \ac{VR}, \ac{AR}, and \ac{MR}) in recent years. \ac{XR} exergames offer a higher degree of enjoyment, motivation, and exercise performance compared to non-immersive ones~\cite{born_exergaming_2019}. According to the recently published scoping review from \citet{karaosmanoglu_born_2024}, current \ac{XR} exergaming research often focuses on promoting physical activity, treating medical conditions, or providing physical training. This aligns with \citet{feltz_exergames_2019}, who summarized the three main objectives of exergames. Despite the projected benefits of social interaction in multiplayer exergaming, little attention has been given to such design in the current \ac{XR} exergaming domain~\cite{karaosmanoglu_born_2024}.

Among various \ac{XR} exergaming platforms, the ExerCube proves to be a desirable platform for designing socially immersive exergames through balancing the immersive gameplay and the social interactions from the co-location~\cite{martin-niedecken_towards_2019}. The ExerCube is a \ac{MR}-based system using HTC VIVE VR trackers to allow users to interact with the surrounding display \rev{(i.e., three cushioned projection walls)} while performing physical activities in a small, trapezoid-shaped space~\cite{martin-niedecken_exercube_2019}. The main benefit of the ExerCube are to reduce the probability of motion sickness during virtual exercise, which is imperative for prolonged, continuous exercising~\cite{faric_what_2019}, \rev{and to provide a safer exergaming environment~\cite{martin-niedecken_exercube_2018}}. Furthermore, user studies suggest that the ExerCube provides similar levels of flow, motivation, and physical and cognitive challenge compared to exercising with a personal trainer~\cite{martin-niedecken_exercube_2019}. A study by \citet{deng_which_2024} compared older adults' perceptions of 2D flat screen, \ac{VR}, and ExerCube-based exergaming by showing them concept illustrations and videos, and found that the ExerCube has a unique potential for facilitating multiplayer joint exergaming that leads to encouragement of social connections and physical activities.

\subsubsection{Cooperative Exergaming}
The social interaction and the bodily interplay within cooperative play are projected as an important contributing factor to positive exergaming experiences~\cite{lee_psychological_2017, mueller_designing_2017}. For example, \citet{marker_better_2015} states that cooperation in exergaming sustains interest and motivation, and promotes long-term motivation, pro-social and helping behaviours, and higher motivation and self-efficacy towards exercising compared to competition. \citet{peng_playing_2013} noted that co-located cooperative exergaming provides a higher enjoyment and future play motivation than single-player modes and remote competitive modes. While remote competitive and single-player modes resulted in greater physical exertion than the co-located cooperative mode, this difference might be attributed to players continuously dividing their attention to avoid colliding with objects or other players in the shared space. \citet{deng_which_2024} also suggested that socially-oriented cooperative exergaming has the potential to increase the adoption rates of exergaming and overall engagement in physical activities. \rev{Similarly, \citet{marquez_segura_design_2013} highlighted the physical and performative social interactions in co-located social exergaming (e.g., making funny gestures and laughing together) that provides increased engagement, arousal, and positive emotions.} Lastly, \citet{martin-niedecken_exercube_2019} find that cooperative modes in co-located exergaming demand more cognitive attention, which results in higher exertion and might be the most balanced in terms of physical exertion and fun. 

Although all of these publications have explored cooperative exergames, they did not look at dedicated cooperative game mechanics but focused on separate interactions or augmenting existing single-player experiences: \citet{peng_playing_2013} and \citet{marker_better_2015} used exergames where players only perform individual actions that have no interaction with one another. \citet{martin-niedecken_towards_2019} used a single-player exergame with added custom rules outside the game (e.g., having one player riding on the back of another). Whereas these previous works provide a theoretical background for our research, they do not provide insights into how cooperative game designs and their resulting social interactions can influence the players' experience and performance. Accordingly, our research addresses this crucial gap by evaluating \rev{the impact of tailored cooperative designs on the player experience and exercise performance} in an immersive cooperative exergaming context. 

\section{Exergame Design: Space Scavenger Squad}
\label{sec:design}
To answer our research questions, we developed a custom two-player exergame in Unity~\cite{unity_technologies_unity_2023} (V2022.3.14f1). We chose the ExerCube as the platform because its \ac{MR}-based concept allows direct social interactions and full-body movements, and its surrounding display leverages immersion and realism for a balanced co-located experience~\cite{martin-niedecken_exercube_2018}. We designed the exergame with the goals of improving social interaction and promoting a healthy lifestyle for players, using full-body movements in a standing position to build a space-themed exergame featuring a reaction-based multiplayer task using the ExerCube. We titled the game \sss. 

The ultimate goal of \sss is to promote physical activity and social interaction in an engaging manner. However, the immediate objective of the game is to assess the selected cooperative designs. Specifically, we aim to evaluate their impact on players' enjoyment, motivation, performance, and cooperation. We limited the complexity of the gameplay \rev{to simple tapping gestures~\cite{stach_classifying_2009}} to allow players more capacity for exertion and cooperation because previous literature has indicated that gameplay and cooperation can interfere with each other~\cite{kaye_exploring_2016}. All game conditions feature the same interaction: on the three panels of the ExerCube, coloured orbs appear, which have to be caught by dashing to that spot and touching the orb physically with either hand. Both players wear positional trackers on both wrists, which capture their movements and allow the game to detect the catching of orbs. To ensure consistency across conditions, we pre-determined a list of coordinates in a counterbalanced distribution so that each of the three panels and each area (upper, centre, lower) of a panel has a similar amount of orbs, and two consecutively spawned orbs are distant enough to encourage movement.

\begin{table*}[t]
\centering
\captionsetup{justification=centering}
\caption{An overview of the cooperation mechanics and implementations of each condition, \\mapped onto the \textit{Forms of Coop} category in \citet{pais_living_2024}.}
\vspace{-1em}
\label{tab:cat}
\begin{tabular}{ll|ccc}
 & & \VOne & \VTwo & \VThree \\ 
\hline
Implementation & Color of Dots & \green{Green} & \blue{Blue}/\orange{Orange} & \purple{Purple} \\
 & Tappable Player & Either & Only Same Color & Either \\
 & Timing of Tap & Individual & Individual & Concurrent \\
 & Spawn Interval & 1.5s & 1.5s & 2.25s$^1$ \\
 & Despawn Time & 3s & 3s & 4.5s$^1$ \\
 & Cooperation & \faStar & \faStar \faStar & \faStar \faStar \faStar \\
\hline
\hline
Arrangement & Free & \green{\faCheck} &  &  \\
 & Strict &  & \faCheck & \faCheck \\
 & Coupled &  & \blue{\faCheck} &  \\
 & Coincident & \faCheck &  & \faCheck \\
\hline
Synchronicity & Sequential &  &  &  \\
 & Concurrent &  &  & \purple{\faCheck} \\
 & Asynchronous & \faCheck & \faCheck &  \\
\hline
Communication & Agnostic &  &  &  \\
 & Limited &  &  &  \\
 & Incentivized &  \faCheck & \faCheck & \faCheck \\
\hline
\multicolumn{5}{l}{\footnotesize $^1$ 50\% longer than the other conditions, determined by pilot testing to maintain difficulty, see \autoref{sec:pilot}.}
\end{tabular}
\vspace{-1em}
\end{table*}

\subsection{Selecting the Cooperative Mechanics}
\label{sec:mechanics}
To decide on the compared cooperative game mechanics, we primarily relied on the framework by \citet{pais_living_2024} while also drawing inspiration from other published exergame designs as well as popular commercial applications. Because one of the main goals of the exergame is to promote cooperative play and social interactions, \rev{we focused on the \textit{Forms of Cooperation} category in this framework to explore how distinct game designs that require different types of interplay may lead to differences in the cooperation and communication between players. Therefore, we designed our three game mechanics deliberately to capture most of the characteristics identified in the framework, while also forming a continuum of cooperation designs from loosely-coupled to closely-coupled~\cite{beznosyk_effect_2012, harris_leveraging_2016}. We note that these mechanics are not mutually exclusive, but share multiple cooperative characteristics. However, each design features one unique mechanic as their key characteristic:  \VOneN cooperation, \VTwoN cooperation, and \VThreeN cooperation (see \autoref{tab:cat}).}

\rev{Given the co-located nature of our exergame, our design choices regarding the \textit{Communication} category were limited. Both players share the same space, see the same information on the three projection walls, and are, of course, free to talk to each other. Also, we aimed to explore how players communicate in such situation and, therefore, decided not to intervene in players' communication explicitly. Instead, we consider playing a bodily game in a shared physical space as \textit{incentivized} by nature since players must take caution to not collide with each other. However, we note that more strongly coupled tasks also increase the communication incentive automatically.}

\rev{In terms of \textit{Synchronicity}, we decided against \textit{sequential cooperation}, which requires player actions to depend on each other's previous actions. Such designs can create undesired downtime or disconnection in gameplay which may disrupt the players' movement flow~\cite{lee_psychological_2017}. However, all remaining options of \textit{Arrangement} and \textit{Synchronicity} are captured by our three cooperative game designs: free (1) vs strict (2), coupled (1) vs coincident (2), concurrent (1) vs asynchronous (2).}

Firstly, \VOneN cooperation provides a loosely-coupled design similar to previous work (e.g., \citet{peng_playing_2013} and \citet{marker_better_2015}). Because it does not demand any cooperation to play, we may observe how players cooperate intrinsically. \rev{This design serves as a baseline, control group measurement of players' experience and performance in our co-located exergame.} Next, \VTwoN cooperation features a more closely-coupled design by assigning players individual tasks that contribute to the shared score. Our implementation in the co-located setting allows us to explore players' spatial coordination and communication when they need to navigate the shared space together. Lastly, \VThreeN cooperation implements a very closely-coupled design by requiring players to rely on each other to perform actions at the same time, allowing us to observe the temporal coordination between players. 

\subsection{Audiovisual Design}
To ensure the generalizability of our findings, we added the following design elements to ensure a positive player experience similar to commercial titles: (1) a narrative~\cite{lu_narrative_2015}, (2) background music and sound effects~\cite{cmentowski_never_2023, faric_what_2019}, and (3) visual effects for feedback~\cite{mueller_exertion_2016}. Therefore, we drew inspiration from a number of existing games (e.g., Sphery Dome~\cite{sphery_ltd_sphery_2023} and Osu!~\cite{herbert_osu_2007} for game mechanic design and implementation, Beat Saber~\cite{jan_ilavsky_beat_2019} for audiovisual design, and Among Us~\cite{innersloth_among_2018} for narrative and theme.)

\begin{figure*}[!ht]
    \centering
    \includegraphics[width=1\textwidth]{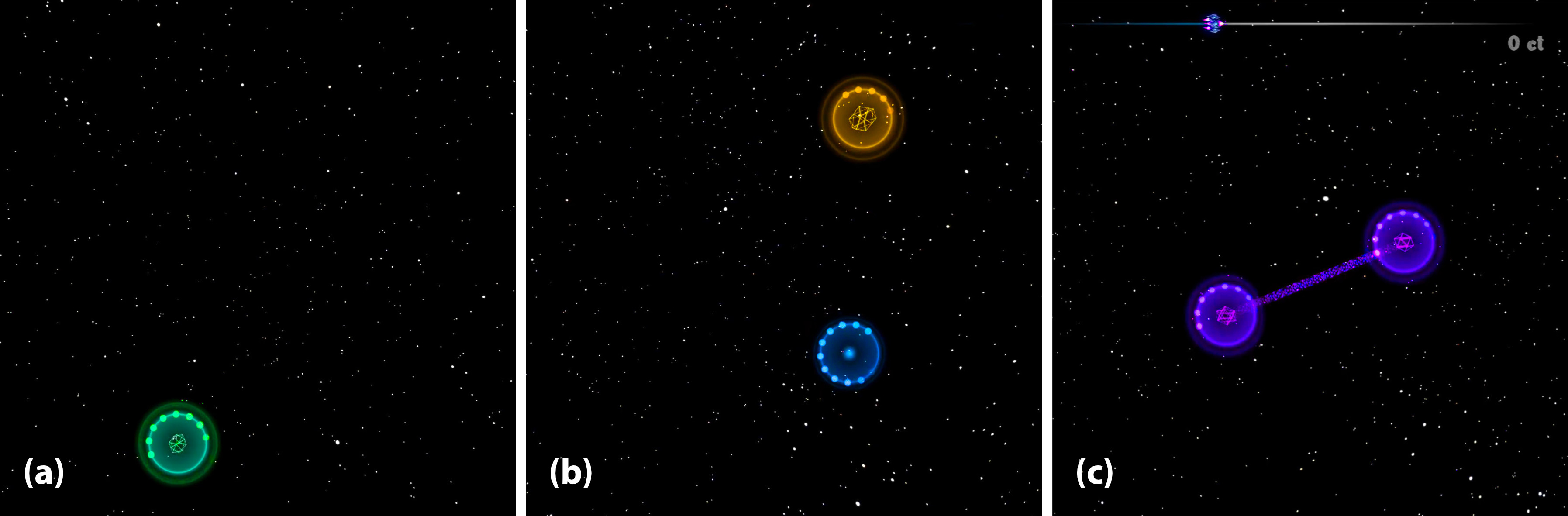}
    \captionsetup{justification=centering}
    \caption{The game interface with (a) \VOne's green orb, (b) \VTwo's blue and orange orbs, (c) \VThree's purple orbs and the HUD showing a progress bar and a score.}
    \Description{This figure provides an overview of the game interface with free cooperation's green orb, coupled cooperation's blue and orange orbs, concurrent cooperation's purple orb dyads, and a Heads-Up Display that shows up on the top of the central panel, including a progress bar and a score.}
    \label{fig:orbs}
    \vspace{-1em}
\end{figure*}

The main HUD of the game includes a score display and a progress bar serving as a timer. The target objective, orbs (see \autoref{fig:orbs}), appear as glowing circles \rev{to elicit interest and enhance engagement~\cite{moreno_socially_2013}}. The outer circle consists of a rotating ring of smaller specks to serve as a timer. Once the ring reaches the full circle, the orb begins to shrink and players will no longer be able to score with the orb once it is barely visible. In the centre of the orbs, we added a pulsating visual effect to make the orbs more noticeable. When a orb is tapped within the time limit, an exploding particle effect plays alongside the display of a score to signify the successful tap.

The game features an outer space theme with subtle and slowly moving particles in a pitch-dark background. We anticipate that this design helps players focus on the orbs rather than the background. During the play sessions, the room's lights are dimmed to ensure the game content is clearly visible.

We used fast-paced background music to foster a sense of urgency. Furthermore, sound effects are played when orbs spawn and score to provide confirmation of players' actions. 

The game has an interactive tutorial (voice-acted by one of the authors of this paper) incorporated with a fictional narrative depicting a space scavenging journey. A narrator first introduces himself as the mission commander, and provides some overall narrative for the game (e.g., there are three ``jobs'' (conditions) today). The tutorial then instructs players to calibrate their ``equipment'' (trackers). At the beginning of each condition, the tutorial provides a bit of narrative context for the condition (e.g., \VTwoN's story happens because a fictional character spilled iced orange soda in space, and the players' job is to collect the blue-coloured ice and orange-coloured soda), as well as the main interaction for the condition. Finally, the tutorial presents a trial version of the orbs for the condition and instructs players to try tapping them. Each player must perform the tapping action successfully once to proceed with the game.

\subsection{Gameplay}
\label{sec:gameplay}
To play the game, players wear two HTC VIVE VR trackers on their two wrists and start standing in the ExerCube. To tap the orbs, players mainly perform stepping movements and arm-reaching motions. Occasionally, jumping and crouching are needed to reach the upper and lower areas of the panels. 

\setcounter{footnote}{1}
At the beginning of the game, players enter a calibration sequence to register their trackers to the game and select their respective colours. Then, players play three conditions of the game in a counterbalanced order. Each game session is three minutes long, plus the interactive tutorial at the beginning. In \VOneN, green orbs spawn every 1.5 seconds and despawn after 3 seconds. Both players can tap every orb and contribute to the shared score. \VTwoN has the same spawn pattern as \VOneN, except orbs appear in two colours: blue and orange. Only the player with the correct colour can collect them. \VThreeN's orbs appear as two simultaneously spawned purple orbs that are connected by a line (see \autoref{fig:orbs}). Players must tap the dyads at the same time (i.e., with a maximum delay of 0.5 seconds). In our pilot testing stage, we learned that group communication requires more time. Therefore, these orbs spawn every 2.25 seconds and despawn after 4.5 seconds to maintain the same perceived difficulty.

\subsection{Game Development \& Pilot Testing}
\label{sec:pilot}
We conducted pilot tests on the prototype with six researchers (in three pairs) from the research lab. After each testing, we readjusted the game and study design to make sure (1) the game has adequate difficulty overall (e.g., we decided on the 3-minute session length; we decided to refer to players based on their assigned colour rather than player number to reduce the complexity of the gameplay), and (2) the three conditions have comparable difficulties (e.g., we decided on the 3-second time limit for \VOneN and \VTwoN, and extended it to 4.5s for \VThreeN). With these adjustments, we aimed at a moderately exerting gameplay that would not be too physically or mentally demanding that it would distract players from the cooperation or their player experience~\cite{martin-niedecken_exercube_2019, ketelhut_integrating_2022}.

\section{User Study}
\label{sec:meth}
To compare how different cooperative designs affect players' enjoyment, motivation, performance, and cooperation, we conducted a within-participants study with three conditions \rev{(baseline and two cooperative designs)}. Our study received ethics approval from the Research Ethics Board (REB) at the University of Waterloo (\#: \anony{46206}).

\subsection{Measures}\label{sec:measures}
We used a demographic questionnaire to collect information about the sample characteristics (i.e., age, gender, ethnicity, experience with cooperative games, experience with exergames, exercise frequency, their current perceived tiredness, and their familiarity with the other participant).

To collect participant's baseline \ac{HR}, we used a Polar H10 \ac{HR} sensor and measured the participants while they were seated and relaxed. We calculated the baseline \ac{HR} by averaging the values recorded for a minimum of five minutes, during which they filled out the demographic questionnaire.

Furthermore, we recorded in-game performance and exertion metrics, participants completed a questionnaire after each condition, and we ended the study with a qualitative semi-structured interview.

\subsubsection{Post-Condition Questionnaires}
After each condition, we administered the following questionnaires:

We used the \ac{PENS} questionnaire~\cite{ryan_motivational_2006} to learn about players' enjoyment and motivation in playing the exergame. This questionnaire has a total of 21 items to be rated on a 7-point Likert scale (\textit{Strongly Disagree/1} --- \textit{Strongly Agree/7}), measuring five subcategories: Competence, Autonomy, Relatedness, Presence/Immersion, and Intuitive Controls. The \ac{PENS} scale is commonly used in exergame and multiplayer XR game studies measuring enjoyment and motivation (e.g., \citet{harris_asymmetry_2019} and \citet{karaosmanoglu_playing_2023}). We preferred the \ac{PENS} questionnaire over other common player experience-related scales such as IMI and PXI because of its focus on the motivational aspect of player experience~\cite{ryan_motivational_2006, ahmadpour_player_2020}.

We also used a \acp{PACES}~\cite{fritsch_study_2022, chen_short_2021}) to monitor participants' enjoyment of the physical activity. This questionnaire has four items, measured with a 5-point Likert scale (\textit{Strongly Disagree/1} --- \textit{Strongly Agree/5}).

To gain more insights into player cooperation, we administered the cooperation subscale of the \ac{CCPIG} V1.1~\cite{hudson_social_2014} measuring subcategories of Team Identification, Social Action, Motivation, and Team Value. This questionnaire has 25 items, offered on a 5-point Likert scale (\textit{Strongly Disagree/1} --- \textit{Strongly Agree/5}).

\subsubsection{In-Game Performance}
Participants' performance in the exergame was evaluated in two parts: game performance and exertion (objective and perceived~\cite{whitehead_exergame_2010}). To measure the game performance, we designed the game to log telemetry data, including each orb's status at despawn (success or fail), the player who tapped it, the time between tap and spawn, and the total arm movement for each condition. This data allows us to calculate players' catch rate, reaction time, and physical movements. Additionally, we recorded participants' \ac{HR} using Polar H10 \ac{HR} trackers. A \ac{RPE}~\cite{borg_borgs_1998} scale (rated from 6 to 20) was also administered after each condition to provide a comparison between perceived exertion and objective \ac{HR}.

\subsubsection{Final Interview}
We conducted one-on-one semi-structured interviews (audio-recorded) after all three conditions of the game with open-ended questions about their experience with the exergame and specific aspects of their cooperation (see supplementary files for the list of questions used in the interviews).

\subsection{Sample Characteristics}
We recruited participants by advertising the study using departmental newsletters and posters on campus. Any participant 18 years or older was eligible to participate if they did not have any physical or mental conditions that prevented them from executing daily physical tasks safely.

Twenty-four participants were recruited for this study (14 female, 10 male). Participants' ages ranged from 19 to 53 years (\mean{25.25}, \sd{8.92} years). Ten participants self-reported as white, 11 as Asian, and three as Indian or South Asian.

Roughly two-thirds (\n{17}) of the participants reported having played cooperative games before. A similar percentage (\n{15}) stated previous experience with exergames (e.g., \textit{Wii Sports} and \textit{Dance Dance Revolution}). However, none of the participants had used the ExerCube before. For existing exercising habits, half of the participants exercise at least weekly (\n{12}), one-third of them exercise bi-weekly (\n{8}), and the rest exercise rarely (\n{6}).

\rev{The interaction between players is as important, if not more so, than the interaction with the game in a co-located exergame~\cite{marquez_segura_design_2013}.} To ensure smoother interaction between players, while eliminating the potential confounding factor of familiarity~\cite{karaosmanoglu_playing_2023}, and to reduce the complexity of scheduling, we asked participants to sign up in pairs with a partner they considered a friend. In the demographic questionnaire, participants rated a 7-point \ac{IoS} scale~\cite{aron_inclusion_1992} to ensure that they are at least familiar with each other (\mean{4.46}, \sd{1.67}).

\subsection{Procedure}
\label{sec:procedure}
After learning about the goals and objectives of the study and providing informed consent electronically, participants put on a Polar \ac{HR} sensor. After checking that the sensor captured the participants' HR, we asked them to sit quietly for five minutes to record a baseline reading of their resting \ac{HR} (a standard procedure~\cite{mandryk_biometrics_2016,  nacke_introduction_2018, karaosmanoglu_feels_2021}). During this time, participants completed the demographic questionnaire and the pre-game measurement of the \ac{RPE} on their own. Then, they were instructed to put on the HTC VIVE trackers, and we introduced them to the ExerCube. 

The game first prompted participants to choose between the colours of blue and orange to represent themselves in the game and calibrate their trackers. Participants then played all three conditions of the exergame in a pre-determined, counterbalanced order to reduce the learning effect. After each condition, participants completed the player experience questionnaires and took a break until they felt ready for the next condition. After all three conditions, two researchers each took one participant to a separate interview room and conducted one-on-one semi-structured interviews. The duration of the study was 53 minutes on average. Participants were remunerated with \anony{\$15 CAD}.

\subsection{Analysis}
\label{sec:analysis}
To answer our research questions, we used a mixed-method approach. With our quantitative analysis, we answer the question of what and how the relevant constructs are affected by the different cooperative designs, and with our qualitative analysis, we further understand the reasons behind their behaviours.

\subsubsection{Quantitative Analysis}
\label{sec:quan_anal}
For quantitative data, we first confirmed the normality of our dependent variables using Shapiro-Wilk tests. When the normality was violated, we used Friedman's ANOVAs instead of one-way repeated measures ANOVAs. For most questionnaires and the in-game metrics, we compared three measurement points: \VOneN, \VTwoN, and \VThreeN. For the \ac{HR} and \ac{RPE}, we additionally compared the conditions against the baseline measurement. We noted that the \ac{HR} readings of the three participants were incomplete because their sensors lost contact during one of the conditions. Accordingly, we discarded these sets and only analyzed the remaining twenty-one sets of \ac{HR} readings. If the ANOVA tests indicated statistical significance, we conducted pairwise comparisons, adjusting the p-values using Holm–Bonferroni corrections. 

\subsubsection{Qualitative Analysis}\label{sec:qual_anal}
To prepare the interview data for analysis, the first author used Dovetail\footnote{https://dovetail.com/} to transcribe the audio recordings, then checked and corrected any inaccurate transcriptions and removed filler words. All researchers then familiarized themselves with the data. To analyze the data, we conducted a thematic analysis (TA)~\cite{braun_thematic_2021} which involved aspects of both reflexive and codebook approaches. Specifically, we consider our iterative process as a reflexive element, while involving multiple researchers coding data independently as an element of a codebook approach. In an iterative process, two researchers individually coded a group of three interviews at a time to find all segments relevant to our \acp{RQ}, using descriptive codes (e.g., ``teammate performance encourages better performance for self'', ``need spatial awareness both for game and for teammate'', and ``friends are quick to communicate and understand each other''). After each group, we met with a tie-breaker to discuss discrepancies in our coding and understanding of the data. This process led to the creation, refinement, and collation of codes. Following the third iteration, the researchers met and developed some initial themes by creating an affinity map from the codes accompanied by participant quotes. We then revisited the already coded data to make sure no detail was lost or misinterpreted in this process. We repeated this process until all data had been coded and reviewed, then refined the themes.

\subsubsection{Positionality Statement}\label{sec:position_statement}
The lead researcher, who designed the exergame, conducted the study and interviews, and led the qualitative analysis, has worked in multiple games user research projects that involved qualitative data collection and analysis. Two senior researchers assisted them in the interview phases, during the qualitative analysis, and the write-up of the resulting themes. Both researchers are proficient in applying qualitative and quantitative methodologies and developing exergame interventions, and have conducted multiple studies on exergame-related topics. During the data analysis process, a third tie-breaker who was not involved in the coding or data analysis process participated in the meetings to balance between insider familiarity and outsider objectivity. This tie-breaker also has experience conducting interview-based data collection and analysis. Lastly, we consulted the other co-authors for clarifications when participants mentioned ambiguous terminologies that could lead to assumptions during the data analysis (e.g., ``immersion'' and ``state of flow''). Taken together, we are confident that this approach allowed us to identify the most relevant and valuable information from the data while staying true to them.

\section{Results}
\label{sec:results}
\subsection{Quantitative Findings}
We first report the results of the quantitative data in our study. See \autoref{tab:questionnaires} and \autoref{tab:hr} for all results and descriptive values. For readability, we only report significant results in text here.

\setlength{\tabcolsep}{2pt}
\begin{table*}[!ht]
\captionsetup{justification=centering}
\caption{Mean scores, standard deviations, and statistical test values of the PENS, the PACES-S, the CCPIG, and performance measures.}
\Description{Mean scores, standard deviations, and statistical test values of the PENS, the PACES-S, the CCPIG, and performance measures.}
\label{tab:questionnaires}

\resizebox{\textwidth}{!}{
\begin{tabular}{p{4cm}p{2cm}p{2cm}p{2.4cm}p{3cm}p{1.4cm}p{1.4cm}}
    \toprule
      & \VOne & \VTwo & \VThree & \textsc{Statistical} & \textsc{\emph{P}} & \textsc{Effect} \\
     & \textit{M (SD)} & \textit{M (SD)} & \textit{M (SD)} &\textsc{Test} &  \textsc{Value}  & \textsc{Size}\\
    \midrule
    \emph{PENS (scale: 1 - 7)}\\
    \hspace{2em}\xspace  Competence & 6.42 (0.56) & 6.28 (0.62)& 5.96 (0.99) & \Fman{2}{3.69} & .158 & \kend{.077}\\
    \hspace{2em}\xspace Autonomy & 5.42 (1.21) & 5.42 (1.24) & 5.46 (1.18) & \F{2,46}{0.02} & .980 & \kend{.001}\\
    \hspace{2em}\xspace Relatedness & 4.97 (1.45) & 5.88 (0.97)& 5.89 (0.99) & \Fman{2}{6.16} & \textbf{.046*} & \kend{.128}\\
    \hspace{2em}\xspace Presence/Immersion & 4.70 (0.93) & 4.80 (0.88) & 4.92 (1.01) & \F{2,46}{1.40} & .256 & \kend{.058}\\
    \hspace{2em}\xspace Intuitive Controls & 6.46 (0.66) & 6.53 (0.47) & 6.19 (0.85) & \Fman{2}{2.06} & .358 & \kend{.043}\\
    \dashedline{7}{54}
    \emph{PACES-S (scale: 1 - 5)}\\
    \hspace{2em}\xspace PACES-S  & 4.74 (0.45) & 4.81 (0.35)& 4.71 (0.56) & \Fman{2}{2.18} & .337 & \kend{.045}\\
    \dashedline{7}{54}
   \emph{CCPIG (scale: 1 - 5)}\\
    \hspace{2em}\xspace Team Identification & 4.22 (0.99) & 4.54 (0.71) & 4.76 (0.45) & \Fman{2}{10.36} & \textbf{.006**} & \kend{.216}\\
    \hspace{2em}\xspace Social Action & 4.30 (0.89) & 4.58 (0.65)& 4.63 (0.48) & \Fman{2}{0.65} & .722 & \kend{.014}\\
    \hspace{2em}\xspace Motivation & 4.25 (0.84) & 4.41 (0.85)& 4.60 (0.50) & \Fman{2}{4.26} & .119 & \kend{.089}\\
    \hspace{2em}\xspace Team Value & 4.50 (0.76) & 4.74 (0.57)& 4.79 (0.44) & \Fman{2}{5.59} & .061 & \kend{.117}\\
    \dashedline{7}{54}
   \emph{Performance Measures}\\
    \hspace{2em}\xspace Catch Rate & 0.98 (0.08) & 0.94 (0.05) & 0.72 (0.16) & \F{1.264,29.076}{51.77} & \textbf{<.001**} & \kend{.692}\\
    \hspace{2em}\xspace Reaction Time & 1.68 (0.21) & 2.26 (0.15) & 2.26 (0.44) & \F{1.269,29.180}{37.19} & \textbf{<.001**} & \kend{.618}\\
    \hspace{2em}\xspace Arm Movements & 316.54 (53.46) & 429.03 (60.89) & 382.87 (64.12) & \F{2,46}{42.23} & \textbf{<.001**} & \kend{.647}\\
    \bottomrule
    &&&& \multicolumn{3}{r}{*\textit{p} $<.05$, ** \textit{p} $<.01$}\\
\end{tabular}
}
\end{table*}

\subsubsection{Enjoyment and Motivation}
Among the five subcategories of the \ac{PENS} questionnaire, only the Relatedness rating differed significantly between the conditions (\Fman{2}{6.16}, \pequal{.046}, \kend{.128}). Post-hoc tests revealed that participants were significantly more aware of their teammates in \VTwoN (\wil{209.50}, \pequal{.022}) and \VThreeN (\wil{60.50}, \pequal{.038}) than \VOneN. Furthermore, the \acp{PACES} did not differ between conditions.

\subsubsection{Cooperation}
Most subcategories of the \ac{CCPIG} questionnaire showed no significant difference between the conditions. However, we found a significant difference in the Team Identification subcategory (\Fman{2}{10.36}, \pequal{.006}, \kend{.216}). Post-hoc tests revealed that participants were more aware of their teammates and considered them more when performing actions in \VThreeN than both \VOneN (\wil{30.00}, \pequal{.049}) and \VTwoN (\wil{17.50}, \pequal{.049}).

\setlength{\tabcolsep}{2pt}
\begin{table*}[!ht]
\captionsetup{justification=centering}
\caption{Mean values, standard deviations, and statistical test values of the objective and perceived heart rates.}
\Description{Mean values, standard deviations, and statistical test values of the objective and perceived heart rates.}
\label{tab:hr}
\small
\resizebox{\textwidth}{!}{
\begin{tabular}{p{1.8cm}p{1.8cm}p{1.8cm}p{1.8cm}p{2cm}p{1.8cm}p{1.2cm}p{1.2cm}}
    \toprule
     & \textsc{Baseline}  & \VOne & \VTwo & \VThree & \textsc{Statistical} & \textsc{\emph{P}} & \textsc{Effect} \\
     & \textit{M (SD)} & \textit{M (SD)} & \textit{M (SD)} & \textit{M (SD)} &\textsc{Test} &  \textsc{Value}  & \textsc{Size}\\
    \midrule
    Heart Rate & 77.78 (15.17) & 120.20 (20.68) & 134.34 (21.19)& 130.12 (20.78) & \Fman{3}{47.80} & \textbf{<.001**} & \kend{.759}\\
    Borg RPE & 10.83 (2.16) & 9.87 (2.95) & 12.88 (2.35)& 12.63 (2.75) & \F{3,69}{14.11} & \textbf{<.001**} & \kend{.380}\\
    \bottomrule
    &&&&& \multicolumn{3}{r}{*\textit{p} $<.05$, ** \textit{p} $<.01$}\\
\end{tabular}
}
\vspace{-1em}
\end{table*}

\subsubsection{Heart rate and Perceived Exertion}
We found a significant effect on the \ac{HR} metric between all conditions and the baseline reading (\Fman{3}{47.80}, \psmall, \kend{.759}). Post-hoc tests revealed significant increases in all three conditions compared to the baseline (\wil{0.00}, \psmall for all three conditions). Between conditions, post-hoc tests showed that participants exerted significantly more in both \VTwoN (\wil{213.00}, \pequal{.001}) and \VThreeN (\wil{36.00}, \pequal{.009}) compared to \VOneN. There is no significant difference between \acp{HR} in \VTwoN and \VThreeN.

Similarly, we observed a significant difference between participants' perceived exertion (\F{3,69}{14.11}, \psmall, \effectpar{.38}). Post-hoc tests showed that participants felt more exerted in both \VTwoN (\ttest{23}{4.82}, \psmall) and \VThreeN (\ttest{23}{4.22}, \pequal{.001}) compared to \VOneN. No significant difference between \VTwoN and \VThreeN. Interestingly, we did not find significance between participants' perceived exertion between \VOneN and baseline. Both \VTwoN (\ttest{23}{4.51}, \pequal{.001}) and \VThreeN (\ttest{23}{3.72}, \pequal{.003}) observed a significant increase compared to baseline.

\subsubsection{In-Game Performance}
There is a significant difference in the catch rate among the three conditions (\F{1.264,29.076}{51.77} (Greenhouse-Geisser corrected), \psmall, \effectpar{.692}). Post-hoc tests revealed that participants performed better in \VOneN than \VTwoN (\ttest{23}{2.96}, \pequal{.007}) and \VThreeN (\ttest{23}{7.49}, \psmall), and better in \VTwoN than in \VThreeN (\ttest{23}{7.56}, \psmall).

We observed a significant difference in participants' reaction time among the three conditions (\F{1.269,29.180}{37.19} (Greenhouse-Geisser corrected), \psmall, \effectpar{.618}). Post-hoc tests revealed that participants took significantly longer to catch orbs in \VTwoN (\ttest{23}{15.29}, \psmall) and \VThreeN (\ttest{23}{6.33}, \psmall) compared to \VOneN. There was no significant difference between \VTwoN and \VThreeN.

Lastly, we also found significant differences in arm movements among the three conditions (\F{2,46}{42.23}, \psmall, \effectpar{.647}). Post-hoc tests revealed that participants moved their arms more in \VTwoN than both \VOneN (\ttest{23}{10.48}, \psmall) and \VThreeN (\ttest{23}{3.58}, \pequal{.002}), and in \VThreeN than \VOneN (\ttest{23}{5.04}, \psmall).

\subsubsection{Ranking according to Perceived Enjoyment and Difficulty}
At the end of the questionnaire, we asked participants to rank the three conditions based on their enjoyment and perceived difficulty. 54.2\% (\n{13}) preferred \VTwoN as most enjoyable, followed by 29.2\% (\n{7}) who preferred \VThreeN, and 16.7\% (\n{4}) who enjoyed \VOneN the most. Furthermore, 75\% (\n{18}) believed \VThreeN was the most difficult condition, followed by \VTwoN (20.8\%, \n{5}), and \VOneN (4.2\%, \n{1}). 

\subsection{Qualitative Findings}
After completing our thematic analysis and affinity mapping steps, we ended up with five themes that broadly focus on the players' experience of playing together with a friend, their cooperation strategies, the importance of shared goals and bodily play, and improvement ideas for future versions of the game.

\subsubsection*{\textbf{Theme 1: Playing with others is fun, even more so with friends}}
All participants enjoyed playing \sss, and particularly emphasized that the cooperative multiplayer aspect is \intext{great for socializing}{21}. Some participants considered our exergame \intext{a team-building exercise for work}{5}. Playing our exergame also allowed participants to better understand each other's behaviours outside the daily norms, specifically, \intext{how would [the teammate] act in the scenarios given to them}{18}. This is especially beneficial for environments where teamwork is essential---for example, as university students constantly engage in \intext{months-long [team-based] projects}{23}, playing our exergame could help them strengthen their bonds that lead to better performance at work.

Above all, participants emphasized the aspect of \intext{playing with a friend}{17} which makes exergaming a lot more fun than playing alone. Many participants enjoyed our exergame as a fun bonding activity: \intext{it creates a memory in your brain, and it is something you can always talk about [in the future]}{17}. As friends know each other's thought process from spending time together, they approach the game with relative ease: \intext{I already knew that we are seeing this [game] similarly}{8}. When performing actions in our exergame, friends can easily \intext{guess what [the other player] is trying to do}{9} and plan their movements accordingly.

While participants were not against trying \sss out with strangers, they foresaw themselves cooperating differently with them and would not \intext{feel as comfortable [as playing with friends]}{17}. Specifically, because participants do not know the other player, they feel the need to \intext{confirm and [prevent] miscommunication}{8} so that their actions do not \intext{cause trouble for [the stranger]}{9}. This means participants would be more reserved in their actions. For example, although \p{23} usually considered themselves as a leader in a team, they would ``\textit{wait to see how the team dynamic plays out}'' to determine if they would lead or ``\textit{the other person [could] call the shots [...].}''

\subsubsection*{\textbf{Theme 2: \rev{Cooperation is organic, competition is inherent}}}
\rev{When discussing their preferred social interactions within the exergame,} most participants \rev{preferred to have} \intext{a cooperative game where you work with a teammate in order to accomplish the goal}{5}. The participants mainly attributed the advantages of cooperative exergaming to two factors. The first factor is the shared emotions when playing cooperatively. When participants reached a high score, they felt a \intext{shared sense of achievement}{10}. This is true even when failing, as \p{8} commented that failing together ``\textit{made succeeding more enjoyable},'' which makes the game feel ``\textit{more fulfilling}.'' One participant also emphasized that the fun in the cooperation comes from the fact that \intext{you can talk [to each other] and you can laugh at the stuff you fail at}{14}. Participants reported that the enjoyment of the cooperative team play motivated them to perform better, and seeing improvements, in turn, increased their enjoyment: \intext{making a higher score made me feel good}{15}. Achievements like these were thought to be important for lasting motivation: they \intext{encouraged me [to] continue to move my body around}{9}.

The second factor is the increased motivation from the responsibility participants felt when playing cooperatively towards a shared goal. The responsibility comes from participants wanting to \intext{help [the other player] and be a positive teammate}{15}. Without it, some participants would not be as motivated, and might \intext{probably put half the effort}{8} when playing alone. One participant further highlighted \VTwoN's strength in fostering cooperative interactions between players by having shared goals as well as separate goals, making it \intext{feel like you are two adventurers in the same journey}{4}.

\rev{In a competitive sense, some participants commented that competition is natural and that as long as there is a score, they would want to compare with someone. Even when playing cooperatively and having only a shared score, they might still \intext{compare [it] with the next batch of friends}{5}. With individual scores, some participants saw it as an opportunity to \intext{have [some] friendly banter}{16} from constantly trying to \intext{one-up each other}{6}. However, competition could also lead to toxicity between the players according to some participants' experience with online team-based competitive games. Evidently, this also happened during our exergame, as a participant harboured complaints about their teammate: \intext{I got really [frustrated] and thought, `you move out of my way so I can get to my spot!'}{17}.}

\rev{Alternatively, a few participants suggested to include some indicators for individual performances so that they can compete with themselves}. Participants wanted to see how well they did \intext{personally aside from the team}{1} so that they could challenge themselves further to improve their physicality, and increase their enjoyment when they overcome their \intext{own record a little bit}{3}. Seeing the teammate's performance motivated some participants to work harder so that they keep up with their teammate to show that they are \intext{a useful asset to the team}{18}. Some participants also tried to encourage their teammate when they noticed their teammate was struggling: \intext{I was encouraging my teammate [by saying] `you are doing a great job', because I wanted us to both do well}{2}.

\subsubsection*{\textbf{Theme 3: Different tasks require different strategies}}
Participants formed different strategies to accomplish the tasks in the three conditions of \sss. For \VOneN, many participants non-verbally agreed to \intext{each take a section}{5} of the ExerCube. Playing with a friend makes it easy for participants to understand each other without explicit communication. For the shared centre panel, some participants chose to leave it \intext{open for both to try to go for it}{5}, while some simply decided based on proximity, that the player closer to the orb was responsible for tapping it. Some participants found enjoyment in being able to help their teammate by covering a larger space. For example, \p{12} volunteered to take the back half of the ExerCube to cover more space than their teammate in the front because their ``\textit{long arms allowed me to touch things [further apart] without having to make [much] effort}.'' 

In contrast, these participants were frustrated about being unable to help their teammate in \VTwoN because of the asymmetric design: \intext{I could have just pressed [the orb] beside me to help. But [in \VTwoQ], I could not}{10}. As \VTwoN requires each player to cover the entire space, participants sought to reduce the workload for their teammate by \intext{always scanning the peripheral for [the teammate] and communicating the colour and location for them}{23}. Some participants emphasized \VTwoN's challenge of their spatial awareness when they had to \intext{make sure that when I touched my [orb], I did not stand in my partner's way, [...] like having to think about how your actions affect someone else's actions}{11}.

\VThreeN was considered the most difficult because participants had to be \intext{on sync}{3} to complete the game. The added mental workload made it more exerting and enjoyable at the same time: \intext{if I were to play [\VOneQ] for 20 minutes versus [\VThreeQ] for 10 minutes, I would still feel more tired because I'm using both my body and my brain. [...] We got [a lower score] in [\VThreeQ], but I still feel more fulfilled because I feel like I put more effort in it.}{8}. Some participants commented that verbal communication is more suitable for \VThreeN because they could then focus more on the game. As such, they quickly decided on a ``trigger phrase''---``\textit{1, 2, 3, go!}'' (\p{3} \& \p{13}) or simply ``\textit{go!}'' (\p{18})---to signal each other on the timing of the tap. Evidently, this still sometimes caused some stress for some participants. As they were focused on trying to tap at the same time, they kept questioning themselves: \intext{`do I do it now?' [...] `is it too late?'}{1}.

The main difficulty in communicating in \sss lies within \intext{not knowing exactly how to communicate}{4} and not having \intext{enough time to communicate [the plan] back and forth}{24}. Some participants found this frustrating. Upon reflecting on their gameplay, some participants also noted things they would have done differently after they had sufficient time to think about it: \intext{I should have let [the teammate] know that I would take the higher ones instead of having [the teammate] jump for them.}{18}.

\subsubsection*{\textbf{Theme 4: Bodily play is important}}
Some participants compared their \sss with other exergames they had previously played. The main appeal of our exergame lies within the physical interactions (i.e., \intext{freedom to move around the [space]}{9} and \intext{hitting [real surfaces]}{17}) that give a tactile feedback. Participants considered the game \intext{an extension of real life}{4} because the interactions with the game were straightforward, reminiscent of real-life sports and exercises (e.g., volleyball (\p{4}), tennis or badminton (\p{8}, \p{16}, and \p{21}), and fencing (\p{8})). 

The \ac{MR}-based design allows participants to clearly see the teammate beside them during the game. Some participants emphasized that being able to see their teammate instead of an avatar will make them less likely to \intext{ignore [the teammate]}{9}, creating a strong team presence. Some participants with previous experience with \ac{VR} gaming commented that they would not have played the game if it was on \ac{VR} because of the \intext{hazards if two players were wearing headsets and can't see each other}{19}.

Physical collision is inevitable in bodily play. While some participants were worried about potentially losing balance and falling over by \intext{running too many steps [rapidly]}{10}, most participants were not concerned about it. Aside from the fact that physical collision is very common in team-based sports, some participants considered crashing into friends \intext{a good laughing factor}{14}. Furthermore, a few participants viewed physical collision as an added obstacle in the game's challenge, that it does not necessarily take the enjoyment away from the game, but can be organically incorporated into it: \intext{[I enjoyed] the moving-around part. I [enjoyed] the anticipation of where the next [orb] will be: [can] I get to it in time? [Can I] not bump into my partner? I feel like that was all a part of the game.}{14}.

\subsubsection*{\textbf{Theme 5: Come for the simplicity, stay for the complexity}}
Overall, we received a lot of feedback for further refining our game design. Some participants commented that the simple mechanic design of the game could provide a lower barrier of entry but may not be able to sustain interest for long-term play. To increase the motivation for future play, participants suggested designing a more complex gameplay. This could be done by \intext{combining different [mechanics] together, so that [the players] need to multitask to both manage their own tasks and to cooperate with the other}{3}. Alternatively, one participant also proposed separating hands and adding leg movements: \intext{I would like that one hand has a colour and the other has another colour, [...] and maybe even tapping with a foot}{1}.

Similarly, some participants noted that the narrative helped to sustain the motivation to play the game by \intext{connecting [them] with the game and feeling like I was actually in the game [...] and giving reasoning as to why we are playing the game and why we are doing certain actions}{1}. To motivate long-term play, some participants suggested including a longer and more diverse \intext{overarching storyline that will trick you into the game [from an] emotional standpoint}{5}.

While most participants approached \sss with expectations from \intext{a gaming perspective}{5}, some saw the potential for the exergame to enhance and even replace traditional physical exercises: \intext{I would be more inclined to play [the exergame] than if I knew it was a workout}{6}. 

Some participants perceived the level of physical exertion from our exergame to be comparable to \intext{cardio exercises, like running on a treadmill}{8}. However, because different participants had different exercise needs and goals, it would be difficult for one game setting to fit everyone. Therefore, some participants suggested having either the option to \intext{control and customize the intensity level of the game}{17}, or an adaptive design that \intext{learns how you are doing and [...] have the game base on your [current and goal] heart rates}{24}. One participant proposed to include an \intext{unlimited mode}{6} where the game runs indefinitely and gradually becomes more difficult by increasing the speed of spawn. This way, the game allows players to try to \intext{meet our limits}{6} and motivate them to perform better in future play.




\section{Discussion}
\label{sec:discuss}
The main goal of our study was to understand how players perceive different cooperation mechanics and how they cooperate in \rev{our co-located exergame}. In this section, we interpret the results from our user study to answer our research questions and discuss implications for future exergame design and research.

\subsection{RQ1: What impact do different cooperation designs have on enjoyment, motivation, and performance in a co-located cooperative exergame?}

Our results show that closely-coupled cooperative designs (\VTwoN and \VThreeN cooperation) incite a higher sense of teamwork and make players consider their teammates more compared to the design that does not require cooperation. Our qualitative results also suggest that players experienced more enjoyment and motivation when playing cooperatively. This extends findings in previous literature (e.g., \citet{peng_playing_2013},\rev{ \citet{stromberg_group_2002}}, and \citet{marker_better_2015}) in showing dedicated cooperative designs provide increased enjoyment, motivation, and performance for exergaming players. Additionally, the feedback from our participants points towards a supporting relationship between enjoyment, motivation, and performance, which aligns with \citet{limperos_understanding_2016}. Interestingly, while our qualitative findings suggest that participants experienced more enjoyment in \rev{the game designs more explicitly requiring cooperation compared to the baseline (\VOneN)}, their physical activity enjoyment (measured with the \acp{PACES} questionnaire) did not differ. Even though this finding might surprise at first glance, it aligns with our game design: all three conditions shared the same type of interactions of movements and only differed in the type of cooperative interaction. As such, this observation \rev{aligns with our qualitative findings} that cooperation makes exergaming more enjoyable and motivating.

\textbf{Themes 1} and \textbf{2} suggest that the social interaction resulting from \rev{cooperative game designs} gives players the motivation to play, enjoy, and perform well in our exergame. Our results show that participants are more aware of their teammates and take them into consideration when performing actions in both \VTwoN and \VThreeN, and both \VTwoN and \VThreeN are more favoured than \VOneN for the cooperation and interaction between players. Some participants preferred \VTwoN because it allowed a sense of autonomy and independence, which gave them the freedom to pursue their own goals. At the same time, it left space for helping each other to reach better shared outcomes. Some preferred \VThreeN for its closer coordination which created a sense of interdependence and responsibility, which made the game more ``meaningful'' for them. Because the goal of \sss is to promote a healthy lifestyle by increasing player's participation in physical exercise, it is important to make sure players are willing to play the game consistently over a long period of time. Previous literature~\cite{feltz_exergames_2019, faric_what_2019} argued that boredom and disinterest in physical activity are some of the main reasons for the lack of exercise. Our qualitative findings suggest that our exergame, after some refinements and improvements, may be able to alleviate this issue. Some participants would not be as interested in playing our exergame in the long term if only by themselves, suggesting the teamwork in the cooperative \rev{designs} provides an additional motivation boost for players to begin the change in exercise habits.

Exergaming in team cooperation helps motivate players to perform well. For players who perform better than their teammates, being able to help their teammate adds to their sense of competence and enjoyment. For the other player, the need to keep up provides the motivation to work hard. Some participants were also motivated to improve their own performance by putting in more effort.

As a game, the exergame inevitably involves some mental workload. Our quantitative data shows that participants in \VOneN perceived a much lower exertion than \VTwoN and \VThreeN, and reacted much faster. Our qualitative findings also suggest that the increased mental workload from the cooperative designs results in higher perceived exertion. This may be attributed to the additional considerations players need to have in \VTwoN and \VThreeN, including interpretation of the player-specific game events, spatial awareness for the game and the teammate, and communication under stress. \rev{To increase the efficiency of communication, players used predetermined ``trigger words'' and deixis to reduce the mental workload in forming and interpreting longer and clearer expressions~\cite{gutwin_descriptive_2002}.}

\textbf{Theme 4} shows that the \ac{MR}-based ExerCube space is an adequate platform for co-located exergames because it allows players to freely move in a clear, visible, and safe space. Our exergame's design resembles real sports or daily activities they are used to performing, increasing the motivation to play. Additionally, this platform also makes it desirable for social exergaming because players can clearly see each other in the space rather than through a virtual avatar.

\subsection{RQ2: How do players cooperate in a co-located cooperative exergame?}
In terms of gameplay, players always intuitively select the ``easiest'' and most direct solutions. In the context of exergaming, this means the most energy-conserving movement. For example, in \VOneN and \VThreeN, most participants unpromptedly divided the space in half to minimize the movement. Even though many participants approached the game knowing that it involved some physical exertion, they did not seek to complicate their movements for the sake of the exercise. The same participants suggested prolonging the game sessions to make the game an alternative for cardio exercises. Additionally, we observed that when \VThreeN occasionally spawned a dyad of orbs within reach of one player, some groups chose to let one participant take both to reduce movement. However, as \textbf{Theme 5} reveals that the complexity of gameplay is essential for long-term engagement in the exergame, designers may need to implement additional rules to gameplay and cooperation to balance player freedom and exertion.

\rev{With \VOneN and \VThreeN, players are allowed a stronger territoriality in that they are able to set up personal spaces by dividing the play space in two. This behaviour coincides with the findings of \citet{azad_territoriality_2012} as the ExerCube provides a display modality equivalent to three vertical large displays. Our participants exhibited the same interaction patterns as findings of this work. Specifically, for out-of-reach orbs, our participants either walked over to tap them, or offered or requested assistance from their teammate. In \VTwoN, however, the territoriality is weakened because players are required to traverse the entire play space, constantly altering the personal space. In this design, while players can play independently, they offered each other help by notifying the teammates of the locations of their orbs through deixis and gestures. Our findings also align with \citet{yang_fostering_2024} in that verbal communication is predominant in co-located immersive play, and that non-verbal cues are also used to assist in communication.} 

\textbf{Theme 3} shows that while participants offer each other help during the games (e.g., keeping an eye for the teammate's orbs and covering the teammate's side in case of need), some participants emphasized that it was because the game's moderate difficulty had left them room to do so. Some participants who considered \VTwoN the most difficult had to stop helping their teammate so that they could focus on their own tasks. Therefore, we speculate that as we increase the physical exertion of the exergame, we may eventually reach a point where individual workload prevents effective cooperation. However, many participants also noted that \VThreeN led to the most overall exertion because of the high mental workload resulting from having to coordinate closely for concurrent actions. This shed some light on a possible solution for this issue: while keeping the physical movement to a moderate, designers can boost the overall exertion by increasing the mental workload from the cooperation. Furthermore, by designing the exergaming experience around effective cooperation, we \rev{immerse players in a cooperative atmosphere that encourages them to engage in cooperation and take advantage of its social benefits~\cite{stromberg_group_2002}.}

Previous literature~\cite{peng_playing_2013} raised the risk of potential collisions as one of the main challenges when designing co-located exergames. \textbf{Theme 4} reveals that this does not necessarily reduce players' enjoyment of the game. To avoid collision, some participants verbally signalled their teammate when they moved so that the teammate was aware of it; one group of participants circled around the space together so that there would be no overlap in their paths; some participants simply laughed and resumed playing when collision nearly happened. Participants explained that because collision is common in team-based sports, they were not worried about it at all. We also speculate that this is because physical collision at the moderate movement speed would not have caused severe injury, which alleviated the participants' concern for it.

While our findings give no evidence pertaining to the impact familiarity may have on the players' experience, players may behave differently \rev{when} cooperating with strangers. \textbf{Theme 1} suggests that friends can communicate and understand each other quickly and intuitively because they know each other's thought processes. This allows friends to predict each other's movements during the game and plan their own movements accordingly. Some participants argued they would be more polite and reserved with strangers (e.g., by communicating to ensure nothing is left to assumptions or misinterpretations). This also applies to potential physical collision, as participants expect themselves to move more carefully to respect the stranger's personal space. 

\subsection{Design Implications \& Recommendations}

Here, we provide three design implications and recommendations that guide future work on developing cooperative exergames.

\subsubsection{Encourage teamwork with autonomy and independence}
Our findings suggest that players favour autonomy and independence even in a cooperative environment. This is evident from the feedback we received for \VTwoN and \VThreeN. Essentially, some players are not inherently cooperative but still enjoy social play, and the independent tasks in \VTwoN gave them a balanced enjoyment between personal challenges and teamwork. We speculate that this consideration may be necessary for pairing up players with noticeably different physical capabilities~\cite{deng_which_2024} or gameplay styles~\cite{tondello_i_2019}. Specifically, we recommend that designers implement both shared and individual tasks in the exergame to provide a variety of gameplay that both increases the game's playability and adaptability for different players.

\subsubsection{Redirect competition to enhance cooperation}
Similar to the above, some players that prefer competition would still play a role in a cooperative environment. While some participants suggested having a team-based competition by using a leaderboard, other participants were worried about the potential toxicity said competition may cause (i.e., some participants complained about their teammate being unable to keep up with them). Therefore, we suggest keeping some competitive elements in the game and using them to fuel cooperation. One possible implementation could be self-competition. We satisfy their need for competition by directing them to compete with their past selves rather than their teammate, which can be done by showing performance scores that reflect individual players' performances.

\subsubsection{Redirect the mental workload to team play}
Finally, our qualitative findings suggest that \VThreeN had the highest mental workload compared to \VTwoN despite the latter showing a marginally higher physical exertion. We believe the reason for this is the coordination required in \VThreeN, which adds extra mental stress for players. As such, exergame designers need to consider balancing mental and physical exertion in the game. Rather than reducing the mental workload, we recommend redirecting it to the teamplay to create not only an effective exergaming experience, but also an enjoyable team-building activity.

\subsection{Limitation}
We acknowledge a number of limitations this study had: (1) The ExerCube is physically located in the \anony{Stratford campus of the University of Waterloo, in a small town primarily consisting of students and older adults}. Therefore, our sample was limited to undergraduate students and University staff. (2) To test the impact of cooperative designs, we kept the game's difficulty to a moderate level. Accordingly, the physical exertion was not comparable to existing HIIT-based training exergames (e.g., \citet{martin-niedecken_hiit_2020}). In light of this, our exergame---in its current state---may not be suitable to replace physical exertion or trainer-led physical training. (3) A multiplayer exergame must consider the physical capabilities of both players. Because players play in the same game space, presenting a fixed-level design may inevitably lead to different exertion levels. We made a few inconclusive attempts during the design process of \sss. We considered adjusting the spawn areas to each player's heights and arm lengths, but it was possible for a taller player to move to the ``shorter'' area, which could lead to an even larger discrepancy in difficulty for \VOneN and \VThreeN. For \VTwoN, we were also worried that this might lead to player dissatisfaction if one player notices the other player getting an easier task than them. Therefore, we unfortunately do not have a solution for this at this stage.

\subsection{Future Work}
In this study, we focused on players' cooperation, and selected three types of mechanics specifically from the \textit{forms of coop} category in \citet{pais_living_2024}. Mechanics from other categories may also be suitable for exergames, and we encourage future work to explore alternative mechanics and implementations and their ability to provide enjoyable, motivating, and effective exergaming experiences. \rev{Similarly, our design limited the players' interaction with the game to a single type---tapping. We suggest future work to expand on alternative interactions previous literature (e.g., \citet{harris_leveraging_2016} and \citet{stach_classifying_2009}) has identified.}

Our findings also suggest that cooperative exergames can be potentially used as a team-building activity. Our exergame presented scenarios that demand effective teamwork and conflict resolution under time pressure. Through these challenges, participants learn about each other's cooperative styles. With this in mind, this experience prepares players for teamwork situations outside the game. Because teamwork has become increasingly important in modern society, we implore future research to examine this more closely. 

\section{Conclusion}
\label{sec:conclusion}
Our research makes a strong case for including social cooperation into exergaming \rev{through tailored cooperative game designs}. Through the design and evaluation of a co-located exergame \sss, we have demonstrated how social interaction and teamwork substantially increase enjoyment, motivation, and exertion performance \rev{compared to designs where social interaction is optional}. Cooperative mechanics promote a sense of shared purpose and camaraderie. These create tangible and intangible benefits for players. We observed a complex pattern of social dynamics, revealed by varied cooperative behaviours. To provide a compelling social exergame experience, cooperative exergames should (a) encourage teamwork with autonomy and independence, (b) redirect competition to enhance cooperation, (c) redirect the mental workload of players to team play. We believe these insights serve as a game design guide for the development of physically engaging and socially enriching exergames.

\begin{acks}
This study was supported by the SSHRC INSIGHT Grant (grant number:435-2022-0476), NSERC Discovery Grant (grant number: RGPIN-2023-03705), CIHR AAL EXERGETIC operating grant (number: 02237-000), the CFI John R. Evans Leaders Fund or CFI JELF (grant number: 41844), the Mitacs Accelerate Grant (grant number: IT40801), the Lupina Foundation Postdoctoral Research Fellowship, and the Provost's Program for Interdisciplinary Postdoctoral Scholars at the University of Waterloo.
\end{acks}

\bibliographystyle{ACM-Reference-Format}
\bibliography{references}

\end{document}